\documentclass[twocolumn,showpacs,preprintnumbers,amsmath,amssymb]{revtex4}
\usepackage{latexsym}

\usepackage{amsfonts}  
\usepackage{amssymb}
\usepackage{graphicx}   

\begin{document}

\title{Vibration-induced climbing of drops}

\author{P. Brunet}
 \email{p.brunet@bristol.ac.uk}
 \author{J. Eggers }
 \author{R.D. Deegan}

 \affiliation{Department of Mathematics, University of Bristol,
     University Walk BS8 1TW Bristol, United Kingdom.}

\date{\today}

\begin{abstract}

We report an experimental study of liquid drops moving against
gravity, when placed on a vertically vibrating inclined plate, which
is partially wet by the drop. Frequency of vibrations ranges from 30
to 200 Hz, and above a threshold in vibration acceleration, drops
experience an upward motion. We attribute this surprising motion to
the deformations of the drop, as a consequence of an up/down
symmetry-breaking induced by the presence of the substrate. We
relate the direction of motion to contact angle measurements. This
phenomenon can be used to move a drop along an arbitrary path in a
plane, without special surface treatments or localized forcing.
\end{abstract}

\maketitle

A drop of liquid on an inclined substrate will slide downward
due to gravity, unless the drop is pinned by contact angle
hysteresis \cite{Dussan83,degennes}. Since the contact angle hysteresis
is reduced by vertical vibrations \cite{Andrieu,DeckerGaroff},
one might expect that sufficiently strong shaking will always
make the drop come loose and provoke it to slide. Here we report
for the first time that on the contrary, sufficiently strong
harmonic shaking in the vertical direction will always cause
the drop to climb up the slope, regardless of system parameters.

\begin{figure}[h!]
\begin{center}
\resizebox{\columnwidth}{!}{\includegraphics{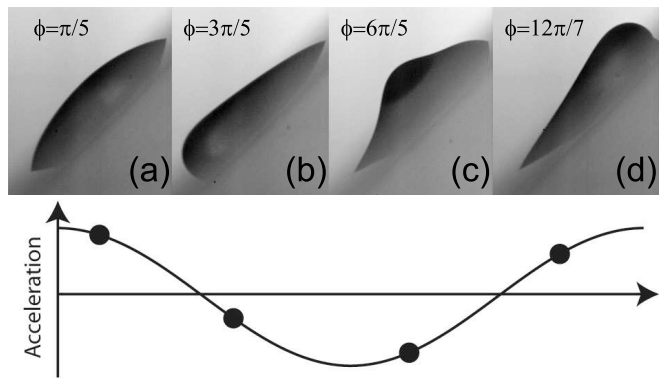}}
\caption{Side view of a climbing drop (and its reflection) on a
vibrating plate inclined at $\alpha$=45$^{\circ}$ at different
phases $\phi$ of the cycle. Images (b) \& (d) show the maximum lower and
upper contact angle, respectively.  Parameters are $V$=5 $\mu$l,
$f$=60 Hz ($f/f_0$=1.18), $a/a_0$ = 1.03, $\nu$= 31 mm$^2$/s. The
lower plot shows acceleration versus phase and the corresponding
acceleration for each of the images.} \label{fig:setup}
\end{center}
\end{figure}

We attribute the upward force to a combination of the broken
symmetry caused by the inclination of the substrate with respect to
the applied acceleration and the nonlinear frictional force between
the drop and the substrate. During the downward acceleration phase,
the drop becomes taller and thus more compliant to lateral forcing.
Hence, the maximum value of the contact angle attained on the upper
side (Fig.~\ref{fig:setup}(d)) is greater than the maximum value
attained on the lower side (Fig.~\ref{fig:setup}(b)), and the drop
thus experiences a net upward force~\cite{footnote}. However, for a
purely linear frictional force the net force on the drop would
average to zero over one period; hence some nonlinearity in the
interaction between the drop and the substrate is needed. This key
issue is illustrated by a model calculation below.

In our experiments a drop of a glycerol-water mixture, of volume $V$
between 0.5 and 20~$\mu$l was deposited on a plexiglass substrate
inclined to the horizontal with an angle $\alpha$ up to 85$^o$. The
resulting sessile drop was between 1 and 3 mm in diameter, and
pinned in the absence of shaking. The substrate was oscillated
vertically using an electromagnetic shaker with acceleration up to
50$g$ where $g$ is the acceleration due to gravity, and frequencies
between 30~Hz and 200~Hz. The acceleration was monitored with a
single-axis accelerometer; the acceleration due to unwanted lateral
motion did not exceed 3\% of the vertical acceleration.

The kinematic viscosity $\nu$ of the various mixtures ranged between 31
and 55 mm$^2$/s. For lower viscosities the drop can break up before
the onset of climbing; for higher viscosities, drops move slower and
thus their dynamics is more difficult to access. The surface tension
$\gamma$ was equal to 0.066 N/m, the density $\rho$ at 20$^{\circ}$C
ranged from 1190 kg/m$^3$ for $\nu$= 31 mm$^2$/s to 1210 kg/m$^3$
for $\nu$= 55 mm$^2$/s.  The contact line angle and position were
measured visually using a high speed camera and diffuse
back-lighting. The advancing and receding contact angles were
measured to be $\theta_a$ = 77 $\pm$ 2$^{\circ}$ and $\theta_r$ = 44
$\pm$ 3$^{\circ}$ by inflating or deflating a drop with a syringe
and observing the yielding point of the contact line.

\begin{figure}
\includegraphics[scale=0.40]{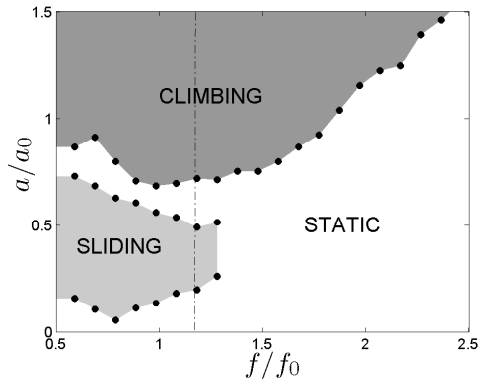}
\caption{Phase diagram of drop motion for $V$=5$~\mu$l,
$\alpha$=45$^o$, and $\nu$=31~mm$^2$/s. The normalization factors
are $f_0 = 50.77$~Hz and $a_0$=174 m/s$^2$.}
\label{fig:phasediagram}
\end{figure}

The acceleration $a = (2 \pi f)^2 \text{A}$, where $A$ and $f$ are
the amplitude and frequency of the applied vibrations, induces a
rocking motion of the drop as shown in Fig.~\ref{fig:setup}. When
the rocking motion is large enough, the contact line begins to
unpin. The resulting mean motion of the drop depends on the values
of $a$ and $f$ as shown in Fig.~\ref{fig:phasediagram}. The drop can
move down the substrate (sliding), remain stationary (static) or
move up the substrate (climbing).

We obtained similar phase diagrams for different drop volumes,
viscosities, and angles of inclination $\alpha$. While the
boundaries of each regime shift as these parameters are varied, the
qualitative appearance of the diagram remains unchanged. All results
reported below are for the parameters of
Fig.~\ref{fig:phasediagram}. As units of frequency and length, we
choose the resonance frequency of the drop's rocking mode $f_0=$50.77~Hz and the linear drop size $V^{1/3}= 1.71$~mm, as this is a measure of the deformation. The value of
the former is calculated from the expression given
in~\cite{Celestini06} (eq. (6)), with $\theta_e = 62^{\circ}$  as
calculated from $\cos\theta_e = (\cos\theta_a + \cos\theta_r)/2$ and
the geometrical parameter $h$=1; our value is similar to that
measured by \cite{Daniel05,Dongetal07}.  We take $a_0= (2 \pi f_0)^2
\text{V}^{1/3}=174$~m/s$^2$ as the characteristic acceleration.

\begin{figure}[h!]
\resizebox{\columnwidth}{!}{\includegraphics[scale=0.40]{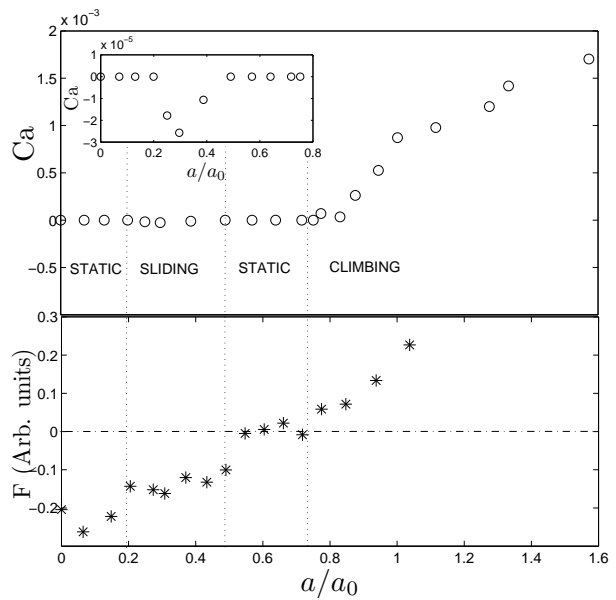}}
\caption{\textit{Top - } The capillary number $Ca$ vs. $a/a_0$
($f$=60 Hz, $V$= 5 $\mu$l, $\nu$ = 31mm$^2$/s), measured along the
dot-dashed line in Fig.\ref{fig:phasediagram}. Inset: magnified view
of data below the climbing threshold. $\text{Ca}=10^{-3}$
corresponds to a speed of 1.79 mm/s. \textit{Bottom - } Estimate of
the force averaged over a cycle due to the difference between the
upper and the lower contact angles.}

\label{fig:speed}
\end{figure}

The drop speed $U$ in terms of the capillary number $\text{Ca}= \rho
\nu U / \gamma$  versus the normalized acceleration for fixed
frequency $f/f_0$=1.18 is plotted in Fig. \ref{fig:speed}(a).
$\text{Ca}=10^{-3}$ corresponds to a speed of 1.79 mm/s. The data
shows that as the acceleration is raised the motion transitions from
static to sliding, back to static, and finally to climbing. In the
sliding phase, speeds are two orders of magnitude smaller than
typical climbing speeds, as shown in the inset. For other parameters
(e.g. $V$=10 $\mu$l, $f$=60Hz or $V$=5 $\mu$l, $f$=45 Hz) the
capillary number threshold for climbing remains the same, but the
threshold for sliding varies significantly.

\begin{figure}[h!]
\begin{center}
\resizebox{\columnwidth}{!}{\includegraphics{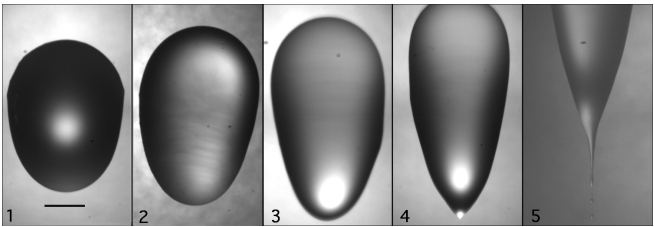}}
\caption{Views from above of sliding (1), static (2), and climbing
drops (3-5). As the speed of the drop increases from (3) to (5), the
trailing end transitions to a corner (4), and then to pearling.
$a/a_0$=0.46 (1), 0.67 (2), 0.96 (3), 1.13 (4), 1.39 (5). The scale
bar equals 1 mm. } \label{fig:above}
\end{center}
\end{figure}

Based on our high-speed observation of the drop viewed from above,
we interpret the progression through the stuck, sliding, and
climbing regimes as follows. For low accelerations the contact line
is pinned at all point on the perimeter.  As the acceleration
increases the pinning progressively breaks due to the rocking motion
until a few points on the side of the drop remain pinned, and only
temporarily (see Fig.~\ref{fig:above}(1) showing sliding).
As the acceleration continues to increase, the upward force due
to shaking cancels out gravity, to produce an almost vanishing
net force. At this point the drop repins along its sides,
and becomes stationary again (see Fig.~\ref{fig:above}(2)).
Yet greater acceleration increases the upward force past the
pinning threshold and the drop moves up the substrate
(see Fig.~\ref{fig:above}(3)-(5)).

\begin{figure}[b]
\includegraphics[width=5cm]{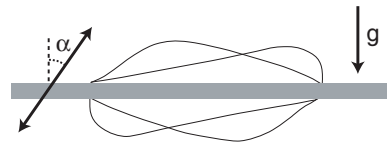}
\caption{Drops resting on a horizontal plate (or hanging below it),
but shaken at an angle. The drop moves in the direction of the shake
pointing normal toward the drop. } \label{fig:horizplate}
\end{figure}

To support the crucial observation that shaking always produces an
upward force, we performed additional experiments using a horizontal
plate subjected to an acceleration angled away from the vertical, thus eliminating gravity.
The results are depicted in Fig.~\ref{fig:horizplate}: sessile drops
move to the right and pendant drops move to the left. This corresponds to there being nothing but climbing motion in the original inclined plate geometry. 

While the drop's contour does not vary significantly over one
period, the contour shape is strongly dependent on the drop's global
velocity.  As shown in Fig.~\ref{fig:above}, the shape for sliding
and climbing drops are similar, but the downslope end tends to
sharpen with increasing Ca. Above $\text{Ca} \approx
1.8\times$10$^{-3}$ the trailing end of the drop undergoes a
pearling instability similar to the observations
of~\cite{Podgorski01,Nolwenn05} for a drop sliding at constant speed
down an incline.

\begin{figure*}
\includegraphics[scale=0.41]{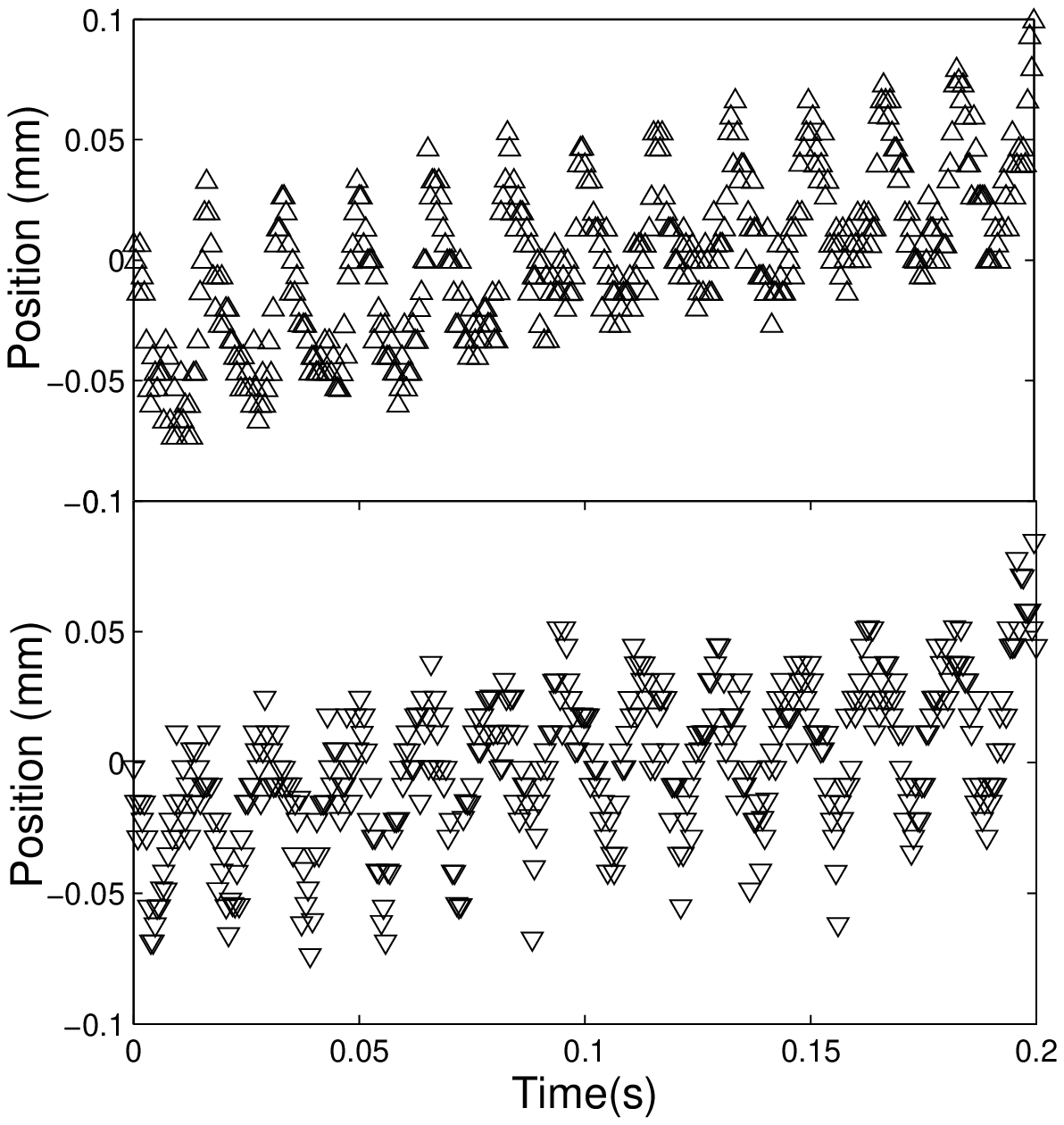}
\includegraphics[scale=0.41]{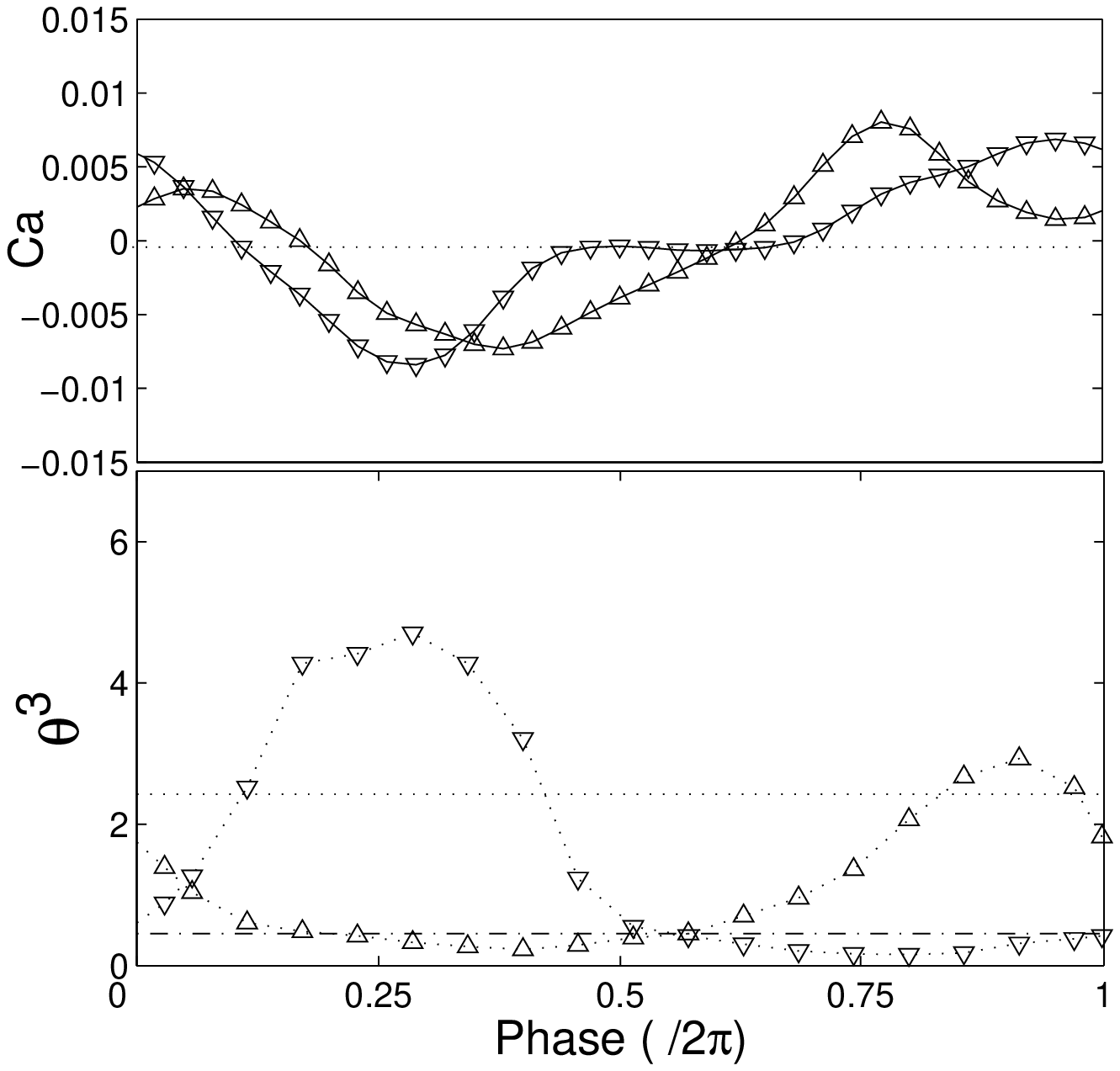}
\includegraphics[scale=0.41]{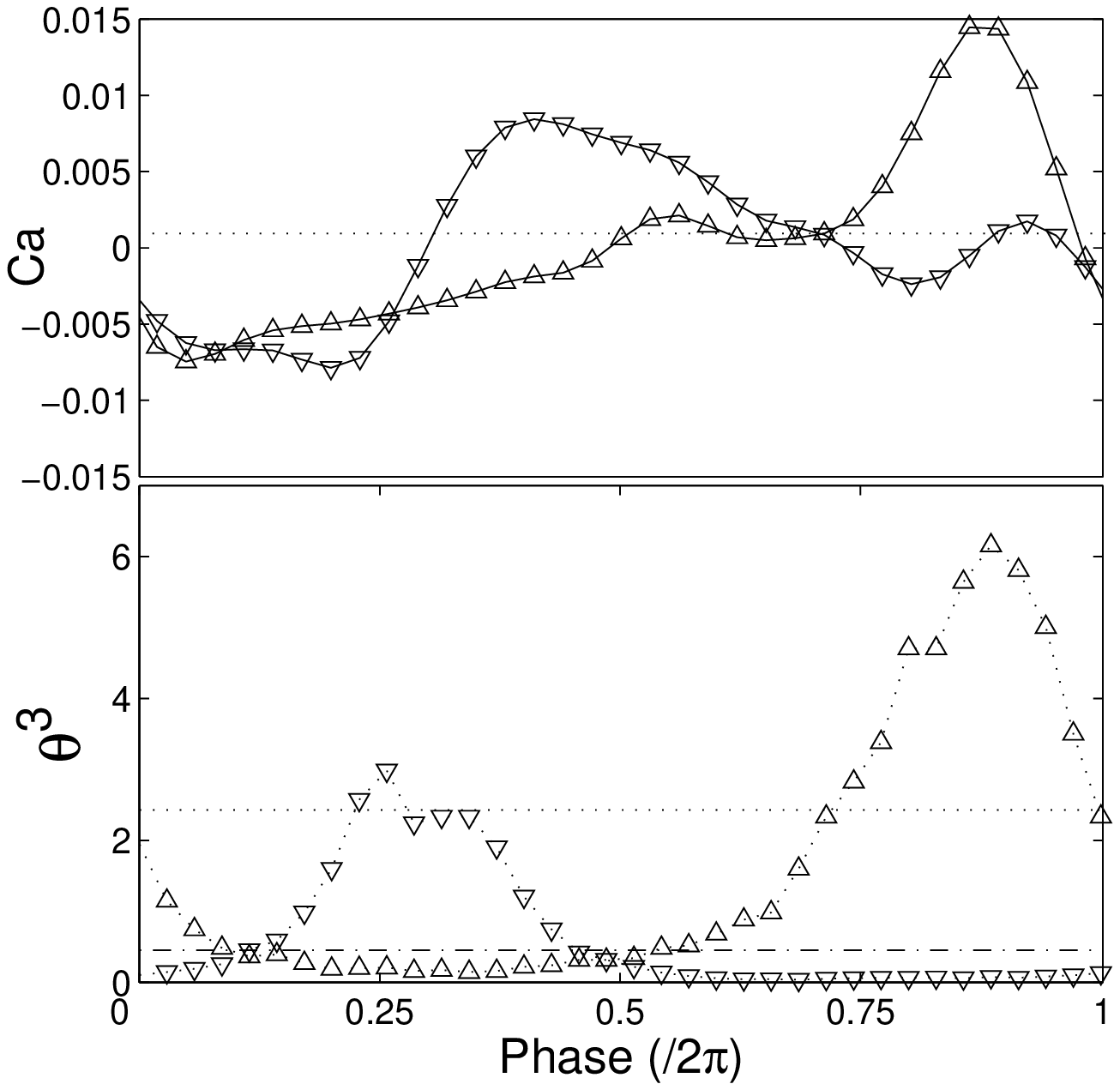}
\caption{\textit{Left Panel }- Displacement of the upper
($\vartriangle$) and lower ($\triangledown$) contact lines for a
climbing drop during 12 periods ($a/a_0$=1.03). Capillary number
(\textit{Top Center \& Right}) and contact angle cubed
(\textit{Bottom Center \& Right}) versus time for one period of
oscillation for a sliding drop (\textit{Center}) and a climbing drop
(\textit{Right}). The average Capillary number is denoted by a
dotted line dashed line in the \textit{Top} plots. $\theta_a^3$ and
$\theta_r^3$ are denoted by dotted and dot-dashed lines,
respectively, in the \textit{Bottom} plots. $f$=60Hz, $V$=5 $\mu$l,
$a/a_0$= 0.45 (\textit{Center}) and 1.03 (\textit{Right}).}
\label{fig:clpos}
\end{figure*}

Figure~\ref{fig:clpos} shows the relationship between the
instantaneous measurements of the contact line's speed and angle for
our drops. The leftmost panel shows an example of raw data for the
contact line position at the upper and lower end for a climbing
drop: oscillations are superimposed on a slow upward motion. The
instantaneous speed was extracted from the position data by
subtracting the mean motion, averaging the result over multiple
periods to reduce noise and fitting it with a spline, and
differentiating the spline. In Fig.~\ref{fig:clpos} the center panel
is for a sliding drop and the right panel is for a climbing drop.

There is a clear correlation between the instantaneous contact angle
and the contact line speed: the maximum speed of the upper and lower
contact line coincides with the maximum upper and lower contact
angle, respectively. Furthermore, the contact angle is a good
indicator of the qualitative difference in the oscillations of a
sliding and climbing drop. For a climbing drop the average of the upper contact angle $\theta_u$  is larger than the average of the lower contact angle $\theta_d$. In contrast, for a sliding drop
the average value of $\theta_d$ for a sliding drop is significantly
greater than that of $\theta_u$. The uphill shift in the asymmetry
of the oscillations with greater acceleration correlates well with
the average motion.

Nonetheless, we find no local and instantaneous law relating the
contact line speed to the angle. In particular, as indicated by the
lack of pinning when the contact angle lies between the advancing
and receding contact angle (denoted by the dotted  and dot-dashed
lines in Fig.~\ref{fig:clpos}), the  Cox-Voinov law \cite{Cox,Voinov} generalized to
account for contact angle hysteresis found for steady drop
motion~\cite{Nolwenn05} does not work. To account for the related
phenomenon of drop motion induced by {\it asymmetric} shaking,
Daniel \textit{et al}~\cite{Daniel05} propose a ratcheting
mechanism.  This mechanism depends crucially on the drop pinning
when the contact angle lies between the advancing and receeding
contact angles.  However, the absence of pinning observed for our
system at elevated accelerations, consistent with the observations
of \cite{Andrieu} and \cite{DeckerGaroff}, contradicts this explanation.

We obtain a simple contact angle criterion for climbing or sliding
from an estimate of the force on the drop, resulting from both the
upward driving force and gravity. The unbalanced Young force per
unit length of the contact line is $\gamma(\cos\theta -
\cos\theta_e)$~\cite{degennes}. Disregarding variations along the
perimeter, the force averaged over a period is \cite{foot2}
\begin{equation}
F = \frac{1}{T} \int^{T}_{0} (\cos\theta_d - \cos\theta_u) dt.
\label{eq:forcelaw}
\end{equation}

The resulting values for the force versus acceleration are plotted
in the bottom graph of Fig. \ref{fig:speed}. A positive value
corresponds to a force directed upwards. The force can be converted
to dimensional units by multiplying by
$V^{1/3}\gamma=1.13\times10^{-4}$ N. The force becomes positive
close to the acceleration threshold for climbing. Below the climbing
threshold this force is an inadequate measure of the speed, due
to our neglect of the retention forces responsible for contact
angle hysteresis. This neglect is only permissible for large
accelarations, for which the hysteresis vanishes
\cite{Andrieu,DeckerGaroff}. Preliminary experiments performed with
much smaller hysteresis (drops of silicone oil on a
partially wetting fluorinated-coating substrate) show much greater sliding speeds, in
accordance with this argument.

\begin{figure}
\includegraphics[width=6cm]{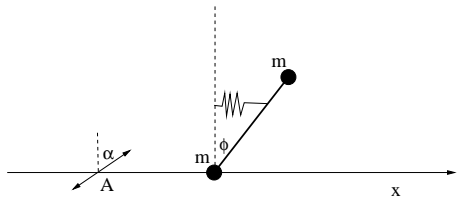}
\caption{A mechanical model for moving drops. The support is
vibrated at an angle $\alpha$ relative to the vertical. In
equilibrium, surface tension holds the drop in a symmetric position,
$\phi = 0$. In the horizontal direction, a nonlinear frictional
force $-F_0 \dot{x}^{\beta}$ acts between the base of the drop and
the support.} \label{model}
\end{figure}

The role of a nonlinear friction law between drop and substrate
is examined through a simple mechanical model which
captures the essence of the drop system. The drop is represented
by a straight rod, held upright by a non-harmonic spring, which
represents surface tension forces (see Fig.~\ref{model}). Thus a
positive angle $\phi$ corresponds to the
drop rocking forward. The lower end of the rod, whose position is
$x(t)$, slides on a one-dimensional rail, which is shaken at an
angle $\alpha$ with amplitude $A$. For fixed $x(t)$, $\phi$ averaged
over one period is nonzero, illustrating the symmetry breaking. We
allow for a frictional force $-F_0 \dot{x}^{\beta}$ between the rod
and its support. Numerical simulations show net motion only if the
frictional force is nonlinear, i.e. $\beta \neq 1$. For $\beta = 1$
the force integrated over one period vanishes, regardless of the
dynamics of the drop itself. For $\beta > 1$, the
rod generically moves to a given direction which reverses if
$\alpha$ changes sign. The net motion arises from a finite $\alpha$
that introduces an asymmetry in the motion $x(t)$ of the rod's base.
Thus, over one period the average force $\overline{F} = -F_0/T
\int_0^T \dot{x}^{\beta}dt$ is nonzero, causing the `drop' to
move.  To summarize the implications of the model, the
only necessary--but crucial--ingredients for droplet motion are a
front-aft symmetry breaking, and a non-linear friction law between
the drop and the plate.

There are two candidate mechanisms to
achieve the latter. First, even though the fluid is Newtonian, the
area of contact between the fluid and the plate is changing in time,
which makes the relationship between the mean drop speed and the
total force non-linear. Second, contact angle hysteresis introduces
non-linearity into the force-speed relationship even for a constant
drop shape. Preliminary comparisons of the above results with those of smaller hysteresis show that climbing speeds at the same accelerations are comparable. This casts doubt
on the necessity of hysteresis for climbing.

We have shown that sessile drops subjected to off-vertical
vibrations experience a net force tangent to the plate in the
direction of the vibrations. Forces greater than gravity can be
easily achieved. The manipulation of sessile droplets is of
increasing importance owing to the advent of microfluidics and the
need to move fluid packets around microfluidic devices.  Our results
suggest a device in which droplets can be moved arbitrarily and in
parallel by independently varying the phase and amplitude of the
vertical and horizontal vibrations for each axis. Recent studies
have demonstrated spontaneous drop motion due to gravity fields
\cite{Podgorski01}, wettability gradients \cite{Thiele06}, an
interplay between thermal effects and ratcheting \cite{Linke06},
asymmetric vibrations \cite{Daniel05}, and chemisorption
\cite{Sumino05}.  By contrast, our transport mechanism would work
for uniform substrates, zero mean forcing, and in the absence of
external imposed gradients.

\textit{Acknowledgments -} We are indebted to J.S. Snoeijer for
useful ideas and stimulating discussions. We thank H.A. Stone for
fruitful discussions.


\begin{thebibliography}{00}
\bibitem{footnote} In experiments, the contact angle lags the
acceleration by $\phi/2\pi\approx 0.1$ which explains the appearance
of the maximum of the upper contact angle in the upperward
acceleration phase, and the maximum of thw lower contact angle in
the downward phase.
\bibitem{foot2}
Although (\ref{eq:forcelaw}) is formally similar to the corresponding
force introduced in \cite{Daniel05}, it differs crucially in that
there is no account of hysteresis.
\end{thebibliography}
\end{document}